\documentclass[conference]{IEEEtran}

\usepackage{dblfloatfix}

\usepackage{cite}
\usepackage{amsmath,amssymb}
\usepackage{graphicx}
\usepackage{booktabs}
\usepackage{multirow}
\usepackage{url}

\title{Utility-Aware Progressive Inference over UDP Packet Blocks for Emergency Communications%
}
\author{
\IEEEauthorblockN{
Jiayue Wang, 
Zhiyuan Ren, 
Tao Zhang, 
Wenchi Cheng}
\IEEEauthorblockA{School of Telecommunications Engineering, Xidian University, Xi'an 710071, China \\
Email: jywang225@stu.xidian.edu.cn, zyren@xidian.edu.cn, zhangtao02@xidian.edu.cn, wccheng@xidian.edu.cn
}
}
\begin{document}

\maketitle

\begin{abstract}
Emergency communications increasingly rely on remote visual inference for timely hazard detection under stringent bandwidth and latency constraints. However, conventional UDP-based visual delivery typically performs inference only after the full payload has been received, even though partially received packet blocks may already contain sufficient task-relevant evidence for reliable decision making. This paper proposes a utility-aware progressive inference framework for emergency communications, which operates directly on UDP packet blocks and determines when sufficient task value has been accumulated for early hazard recognition. Specifically, the sender estimates packet-level decision utility as lightweight control metadata, while the receiver progressively updates partial observations, accumulates the utility of received packets, and triggers an early stop once the normalized utility exceeds a prescribed threshold. Experiments on a fire-scene detection dataset show that, at the main operating point, the proposed method reduces the average packet budget by 34.2\% and the decision delay by 1209.17 ms while retaining 91.5\% of the full-reception match rate. The method also maintains its advantage over the stability-based baseline under moderate packet loss and different packet-arrival orders. These results demonstrate that packet-level utility provides an effective basis for communication-efficient and delay-aware hazard recognition over UDP-based emergency links.
\end{abstract}


\section{Introduction}
Emergency communications for hazard discovery, disaster monitoring, and UAV reconnaissance increasingly rely on timely machine decision loops. In these latency-critical scenarios, the communication objective shifts from flawless image reconstruction to early and reliable threat recognition. Recent task-oriented and semantic communication studies have emphasized prioritizing downstream task value over raw bit fidelity \cite{Sagduyu2023TaskOrientedNextG,Wheeler2023SemanticSurvey,Liu2024SemanticSurvey,Lu2024SemanticsEmpowered,Zhang2025EmbodiedReview}. However, many emergency edge devices cannot support computationally intensive transmitter-side semantic extraction, and standard imagery is therefore often streamed to a remote decision engine. In practical emergency links, such imagery is commonly delivered over UDP to reduce protocol overhead and transmission delay, making partial payload availability under packet loss, out-of-order reception, and jitter a natural operating condition. The key issue considered in this paper is therefore not the reliable recovery of all missing data, but whether continued transmission remains necessary once sufficient task evidence has arrived. A \emph{full-reception-first} workflow delays decision making until the entire payload has been buffered and reconstructed. As a result, useful visual evidence already contained in a partially received payload cannot be exploited in time, delaying emergency response.

To reduce both latency and communication cost, a progressive machine-decision paradigm is needed, in which incomplete reception is treated as an intermediate decision state rather than an error state, and redundant transmission can be terminated once sufficient task evidence has been accumulated. Related efforts have explored progressive feature delivery, task-oriented retransmission, and collaborative inference under constrained links \cite{Shao2022EdgeVideoAnalytics,Lan2022ProgressiveTransmission,Li2024TaskOrientedARQ,Song2024UAVSemanticOD,Liu2025CollaborativePerception}. However, these studies mainly operate at the semantic or feature layer, while the protocol-level problem of packet-arrival-aware stopping for reliable emergency inference remains insufficiently formalized. This paper addresses this missing packet-level perspective by treating protocol-aligned payload blocks as the fundamental units for progressive machine decision and by evaluating the accumulated task value of received blocks to determine whether further reception remains necessary. The central question is therefore: \emph{how can the task relevance of protocol-native packet blocks be quantified, and how can an early stop be triggered once sufficient decision evidence has been accumulated?}

To answer this question, this paper proposes a packet-utility-driven progressive decision framework for latency-critical emergency communications. Rather than waiting for full payload delivery, the task utility of the arriving packet stream is continuously evaluated, and an early machine decision is triggered once sufficient decision evidence has been accumulated. The detector backbone itself is left unchanged, and the proposed logic is introduced as a lightweight communication-aware layer over standard task backbones such as YOLO \cite{Redmon2016YOLO,Ge2021YOLOX}. The goal is not to redesign the visual detector, but to expose packet-level task value to the transport process so that receiver-side stopping can be performed under partial UDP reception. In this way, deployability is preserved, while network packetization, transport dynamics, and delay-aware machine decision are coupled within a unified framework.

The main contributions are summarized as follows:
\begin{itemize}
\item \textbf{Protocol-Native Packet Utility Modeling:} We develop a task-oriented utility model that quantifies the decision value of protocol-aligned packet blocks under partial reception. This model directly links UDP payload units with downstream emergency machine-decision value, enabling packet-level reasoning beyond conventional full-reception pipelines.
\item \textbf{Utility-Aware Progressive Stopping:} We design a receiver-driven stopping mechanism that accumulates packet-level utility from received blocks and terminates transmission once sufficient task evidence has been obtained. Unlike detector-output-only early stopping, the proposed rule operates on protocol-native packet utility and therefore links partial payload reception directly with decision latency and communication overhead.
\end{itemize}

The rest of this paper is organized as follows. Section II reviews related work. Section III presents the system model and problem formulation. Section IV describes packet utility estimation and utility-aware stopping. Section V reports the experimental setup and main results. Section VI concludes the paper.

\section{Related Work}

Existing studies relevant to this work can be summarized from three directions.
First, task-oriented and semantic communication methods emphasize downstream inference quality rather than bit-level reconstruction \cite{Sagduyu2023TaskOrientedNextG,Wheeler2023SemanticSurvey,Liu2024SemanticSurvey,Lu2024SemanticsEmpowered,Zhang2025EmbodiedReview}. These works provide the general motivation for task-value-aware communication and show that communication resources should be evaluated according to their contribution to the final task. However, they usually formulate the problem at the semantic or representation level, where the transmitted units are learned features, semantic messages, or task-specific latent variables. In contrast, the present work considers standard visual payloads carried by UDP packets and studies how their protocol-aligned blocks can support progressive machine decision.

Second, progressive transmission and task-oriented retransmission methods reduce communication cost by delivering task-relevant information incrementally or by allocating retransmissions according to inference value \cite{Lan2022ProgressiveTransmission,Li2024TaskOrientedARQ}. These methods are closely related to the idea that not all transmitted information contributes equally to the task. Nevertheless, their decisions are mainly made at the transmitter side or over learned feature units, and the receiver-side stopping process under an actual packet arrival sequence is not explicitly modeled. This paper instead focuses on the accumulated task value of the packet blocks that have successfully arrived at the receiver, which makes the stopping rule directly coupled with partial reception, packet loss, and feedback delay.

Third, communication-aware visual inference and collaborative perception methods improve perception performance under constrained links through edge analytics, semantic object detection, or cooperative sensing \cite{Shao2022EdgeVideoAnalytics,Song2024UAVSemanticOD,Liu2025CollaborativePerception}. These approaches demonstrate the importance of coupling communication and visual inference, especially when wireless resources are limited. However, many of them abstract the transmission process into feature exchange or high-level semantic delivery, so packet-level partial payload availability, out-of-order arrival, and early termination over incomplete visual data are difficult to characterize directly. For emergency visual inference over UDP, these protocol-level factors can strongly affect both decision quality and response delay.

Different from the above works, this paper treats protocol-aligned packet blocks as the basic units of task reasoning. Packet-level utility is estimated before transmission and accumulated at the receiver, so that early stopping can be triggered once sufficient task value has arrived under partial UDP reception. This design also separates the stopping decision from packet scheduling: the arrival order can be externally given, while the receiver still evaluates the task value of the packets that have actually arrived. Therefore, the proposed framework complements existing semantic transmission and scheduling methods by providing a packet-arrival-aware stopping mechanism at the receiver side.

\section{System Model and Problem Formulation}

This section formalizes the emergency visual delivery process, the progressive inference loop at the receiver, and the resulting stopping objective.

\subsection{Network and Transmission Model}

We consider an emergency communication link in which an image is transmitted from a sender to a remote receiver over a connectionless UDP packet stream. The source image is partitioned into $N$ fixed-size packet blocks,
\begin{equation}
\mathcal{P} = \{p_1, p_2, \ldots, p_N\}.
\end{equation}
These blocks are transmitted over an unreliable forward packet channel. Let $k \in \{1,2,\ldots,N\}$ denote the transmission step. Owing to delay, loss, and out-of-order delivery, only a subset of the transmitted blocks may be available at the receiver after step $k$. The cumulative set of successfully delivered blocks is denoted by
\begin{equation}
\mathcal{R}_k \subseteq \{p_1, p_2, \ldots, p_k\}.
\end{equation}

In addition to the forward path, a lightweight reverse control path is assumed, through which a stop signal can be issued once a satisfactory decision state has been reached. After this signal is received by the sender, the remaining packet blocks are no longer transmitted, thereby reducing the transmitted payload and communication cost.

\subsection{Progressive Inference Model}

At transmission step $k$, a partial observation is constructed from the packet blocks available at the receiver:
\begin{equation}
x_k = \Phi(\mathcal{R}_k),
\end{equation}
where $\Phi(\cdot)$ maps the received packet blocks to their corresponding spatial regions and fills missing regions with placeholder padding. A task model $f(\cdot)$ is then applied to the partial observation:
\begin{equation}
D_k = f(x_k),
\end{equation}
where $D_k$ denotes the detection result obtained at step $k$.

As new packet blocks arrive, the receiver executes a progressive receive-update-infer loop. To support early termination, a stopping statistic $U_k$ is evaluated from the current reception and inference state. Once $U_k$ exceeds a stopping threshold, a stop signal is returned through the reverse control path. Since feedback is not instantaneous, $\Delta_k$ packet blocks may still be transmitted or remain in transit after the stopping decision. The final transmission cost is therefore determined jointly by the stopping step and the feedback delay. The concrete form of $U_k$ is specified in Section IV.

\subsection{Problem Formulation}

At each step $k$, a stopping policy $\pi$ maps the current reception and inference state to one of two actions: stop and output $D_k$, or continue waiting for additional packet blocks. Let $\tau$ denote the resulting stopping step. The communication-task tradeoff is formulated as
\begin{equation}
\min_{\pi} \; \mathbb{E}_{\pi}\!\left[\mathcal{L}_{\text{task}}(D_{\tau}, y) + \lambda_1 B_{\tau} + \lambda_2 T_{\tau}\right],
\label{eq:objective}
\end{equation}
where $y$ is the ground-truth annotation, $\mathcal{L}_{\text{task}}$ is the task loss, and $\lambda_1,\lambda_2 \ge 0$ are weighting factors that balance task performance, communication cost, and decision delay. The expectation is taken over channel randomness, packet-arrival uncertainty, and the stopping behavior induced by $\pi$.

The terms $B_{\tau}$ and $T_{\tau}$ characterize two operational costs in emergency links. Specifically, $B_{\tau}$ denotes the effective delivered payload size accumulated before the sender reacts to the stop signal, whereas $T_{\tau}$ denotes the time to action, including forward packet delivery, local progressive inference, and reverse signaling delay. This formulation couples communication resource consumption and emergency response timeliness in a single objective.

Solving \eqref{eq:objective} exactly is difficult because future packet arrivals are uncertain, the task response to partial observations is nonlinear, and a complex stopping policy may introduce additional computation delay and increase $T_{\tau}$. Therefore, a lightweight protocol-aware packet-utility model is introduced to derive a practical receiver-side stopping rule.

\section{Utility-Aware Progressive Detection}

This section instantiates the stopping formulation introduced in Section III through a lightweight utility-aware rule. The rule is intentionally lightweight so that early stopping itself does not introduce substantial computation delay; its novelty lies in using protocol-native packet utility, rather than detector-output stability alone, as the stopping statistic under partial UDP reception. Rather than solving the policy optimization in \eqref{eq:objective} exactly, a practical stopping statistic is constructed from packet-utility estimation and then used for receiver-side early stopping.

\subsection{Progressive Execution Pipeline}

The proposed method is executed as an event-driven receive-update-infer loop governed by the arriving packet stream. Each protocol-aligned UDP packet block is associated with a fixed spatial region on the receiver-side reconstruction canvas. When new blocks arrive, the current partial observation $x_k$ is updated, and the corresponding task state $D_k=f(x_k)$ is evaluated according to the progressive inference model in Section III. In this way, the decision process is advanced directly by packet accumulation rather than by complete-image reconstruction.

\begin{figure}[t]
\centering
\includegraphics[width=\columnwidth]{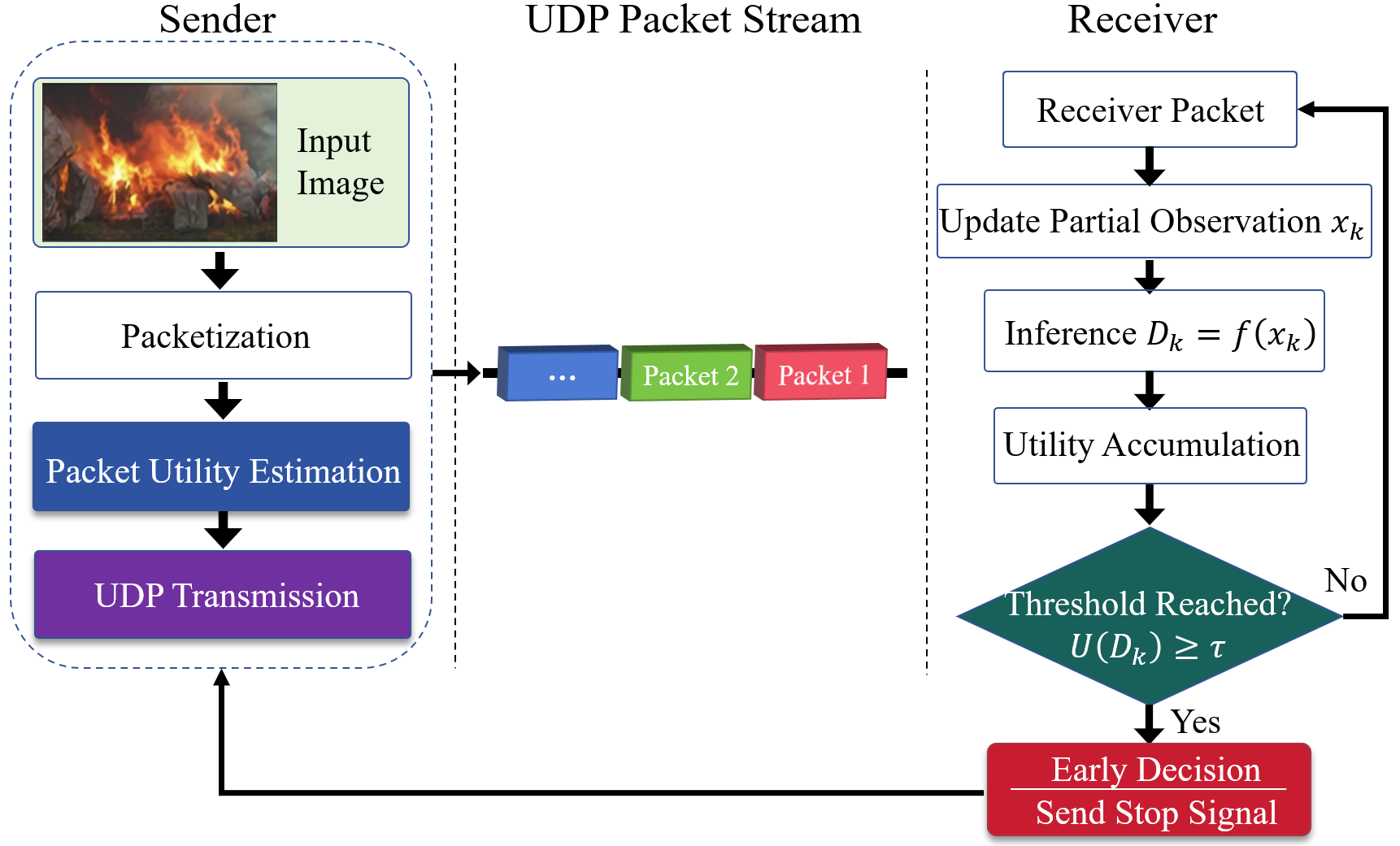}
\caption{Overview of utility-aware progressive machine decision. The sender estimates packet utility before transmission, UDP packet blocks arrive progressively at the receiver, and the receiver updates the partial observation, accumulates utility, and issues a stop signal once the decision threshold is satisfied.}
\label{fig:method}
\end{figure}

The task backbone is left unchanged, while packet-level utility estimation and stopping are introduced on the communication side. As illustrated in Fig.~\ref{fig:method}, packet utility is estimated before transmission, packet blocks are delivered progressively, and a stop signal is issued once sufficient task value has been accumulated at the receiver.

\subsection{Protocol-Native Packet Decision Utility Modeling}

To support receiver-side stopping, the task value carried by each packet block must first be quantified. Let $x$ denote the source image, and let $s(x)$ denote the target detection confidence obtained from the fully received image. For packet block $p_i$, the protocol-native decision utility is defined as the confidence degradation caused by removing that block from the payload:
\begin{equation}
c_i = s(x) - s(x_{\setminus p_i}),
\label{eq:contrib}
\end{equation}
where $x_{\setminus p_i}$ denotes the image reconstructed without packet block $p_i$. A larger $c_i$ indicates that the corresponding packet block carries higher task value.

Equation \eqref{eq:contrib} gives the core utility definition. In practice, the supervision used to train the utility predictor follows the leave-one-block-out procedure in \eqref{eq:contrib}, together with a task-consistent detection matching rule. After each packet block is removed, the fire detector is re-evaluated on the masked image, and the resulting detection is compared with the full-image detection. If the fire target is no longer detected, the matched confidence is set to zero. If the target is still detected but its predicted box has weak spatial overlap with the full-image box, the matched confidence is penalized to reflect spatial inconsistency. The block utility is then computed as the non-negative drop between the full-image confidence and this matched confidence. Applying this procedure to all packet blocks yields a packet-utility map $\mathbf{c}\in\mathbb{R}^N$ for the source image.

Because this empirical utility map is not available during online partial reception, a lightweight CNN-based sender-side regressor $g(x)\rightarrow \hat{\mathbf{c}}$ is trained to predict the packet-utility map from the source image. During deployment, $g(x)$ is evaluated before transmission, and lightweight control metadata is attached to the packet stream, including the predicted utility of each packet block and the total predicted utility of the image. This prediction is not used to rearrange the packet transmission order; instead, it enables the receiver to quantify how much task value has been accumulated under the given arrival process. In this way, real-time stopping can be supported without requiring the receiver to estimate the utility of missing packet blocks online.

\begin{figure}[t]
\centering
\includegraphics[width=\columnwidth]{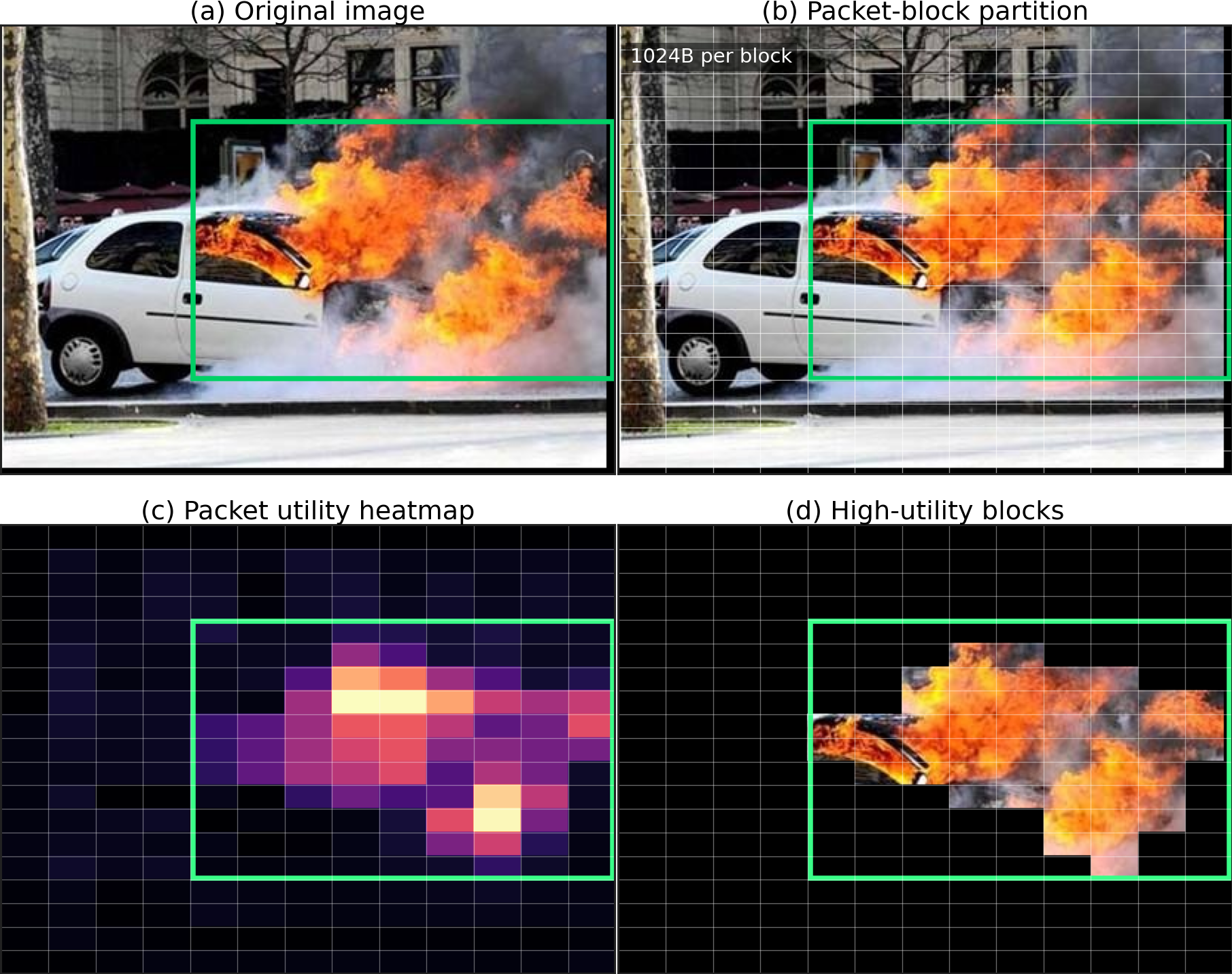}
\caption{Illustration of packet-level utility modeling. (a) Original fire image. (b) Protocol-aligned packet-block partition. (c) Predicted packet utility heatmap. (d) High-utility packet blocks highlighted on the original image. Higher utility values concentrate around task-relevant regions, indicating that different packet blocks contribute unequally to the downstream detection task.}
\label{fig:utilitymap}
\end{figure}

An illustrative example is shown in Fig.~\ref{fig:utilitymap}. The predicted packet utility is concentrated near the fire region rather than being uniformly distributed, which supports the use of cumulative utility as the stopping signal under progressive packet reception. Unlike pixel-level saliency maps, the utility is defined over protocol-aligned packet blocks and reflects their estimated contribution to the downstream decision. The high-utility blocks highlighted in Fig.~\ref{fig:utilitymap}(d) further illustrate the spatial correspondence between the predicted utility and the task-relevant fire region.

\subsection{Utility-Aware Stopping Rule}

At inference time, the stopping statistic introduced in Section III is instantiated using the sender-provided packet-utility metadata. The packet arrival order is assumed to be externally given, and the predicted utility is used only for receiver-side stopping rather than for transmission reordering. Let $\hat{c}_i$ denote the predicted utility of packet block $p_i$, and let
\begin{equation}
C_{\mathrm{tot}}=\sum_{i=1}^{N}\hat{c}_i
\end{equation}
denote the total predicted task utility of the image. During transmission, lightweight control metadata is attached to the packet stream, including the predicted utility of each packet block and the total predicted utility of the image.

At reception event $k$, the receiver has buffered the packet subset $\mathcal{R}_k$. The stopping statistic is defined as the accumulated utility ratio
\begin{equation}
U_k \triangleq \rho_k = \frac{\sum_{p_i \in \mathcal{R}_k}\hat{c}_i}{C_{\mathrm{tot}} + \epsilon},
\label{eq:rho}
\end{equation}
where $\epsilon$ is a small constant for numerical stability. This ratio quantifies the fraction of predicted task value accumulated under the current packet arrival process.

The receiver stops and outputs the current detection result once
\begin{equation}
\rho_k \ge \tau_\rho,
\label{eq:contrib-stop}
\end{equation}
where $\tau_\rho$ is a task-utility threshold chosen according to the target decision requirement. In this way, stopping is tied directly to the fraction of predicted task value that has been received, and transmission is terminated once sufficient utility has been accumulated.

\section{Performance Evaluation}

This section evaluates the proposed framework from the perspective of the communication-task tradeoff achieved by progressive stopping. Since each packet block has a fixed payload size and the simulator uses a fixed inter-arrival interval, the number of received packet blocks directly reflects the delivered payload cost and is also proportional to the idealized transmission delay. Accordingly, the experiments focus on stopping efficiency and task-level decision quality rather than on optimizing the detector itself.

\subsection{Experimental Setup and Baselines}

The proposed framework was evaluated on a fire-scene detection dataset using an offline event-driven packet-arrival simulator. Each image was partitioned into fixed-size 1024B packet blocks, corresponding to spatial blocks of $32 \times 16$ pixels. Unless otherwise stated, packet delivery followed a center-first arrival order with a fixed inter-arrival interval of $5$ ms. Packet-loss robustness was evaluated by randomly dropping packet blocks along the arrival timeline, while missing regions were padded with zeros at the receiver.

A YOLOv8n-based fire detector was used as the task model. It was initialized from a pretrained checkpoint, fine-tuned on partial packet-block observations, and then used to generate packet-utility supervision in a leave-one-block-out manner. For each training image, only samples with a valid full-image detection were retained. Each packet block was removed once, and the detector was re-evaluated on the masked image. The utility target was computed using the task-consistent matching procedure described in Section~IV-B, where the block utility was obtained from the non-negative confidence drop relative to the full-image detection. Thus, the predictor supervision was treated as a task-consistent surrogate of the ideal utility in \eqref{eq:contrib}.

A lightweight CNN-based utility predictor was trained on the resulting contribution maps to estimate the packet-utility map before transmission. The predictor took the full image as input and regressed a block-level utility map matching the packet-block grid. The predicted packet utilities were then attached as lightweight stopping metadata and used by the receiver under a given packet-arrival order.

Three schemes were compared: \emph{Full-Reception Decision}, which performed inference only after all packet blocks had been received; \emph{Stability-Based Early Decision}, which used confidence, class agreement, and box overlap across consecutive detections for stopping; and the proposed \emph{Utility-Aware Early Decision}, which used the stopping rule in \eqref{eq:contrib-stop}. The stability-based scheme serves as a protocol-compatible heuristic baseline that represents detector-output-driven early stopping without packet-level utility information. Performance was reported in terms of matched-ground-truth rate, average received packet blocks, and average decision delay. Packet-loss behavior was further evaluated under different random dropping rates. We use the matched-ground-truth rate as the primary task-quality metric. A test image is counted as successfully matched if the detector output contains at least one fire prediction whose bounding box overlaps with a ground-truth fire box by IoU $\geq 0.5$. This metric serves as a conservative task-level success measure for comparing progressive reception policies, rather than as a full detector mAP benchmark.

\subsection{Main Results}

Table~\ref{tab:e1_main} summarizes the communication-task tradeoff achieved by different stopping policies. The stability-based baseline at threshold $0.9$ achieves a matched-ground-truth rate of $0.6687$ with $473.67$ received blocks and a decision delay of $2368.36$ ms. In comparison, the proposed utility-aware policy achieves matched-ground-truth rates of $0.7741$ and $0.8133$ at $\tau_\rho=0.8$ and $\tau_\rho=0.9$, respectively. These results show that explicitly modeled packet utility provides a more effective stopping signal than heuristic detection stability when decisions are made over a partial UDP packet stream.

\begin{table}[h]
\caption{Main results under progressive stopping policies on the fire-scene dataset.}
\label{tab:e1_main}
\centering
\begin{tabular}{lccc}
\toprule
Policy & Match Rate & Blocks & Delay (ms) \\
\midrule
Full & 0.8464 & 706.29 & 3531.46 \\
Stability-based ($0.9$) & 0.6687 & 473.67 & 2368.36 \\
Utility-aware ($\tau_\rho=0.7$) & 0.7289 & 383.27 & 1916.34 \\
Utility-aware ($\tau_\rho=0.8$) & 0.7741 & 464.46 & 2322.29 \\
Utility-aware ($\tau_\rho=0.9$) & 0.8133 & 561.48 & 2807.38 \\
\bottomrule
\end{tabular}
\end{table}

As shown in Table~\ref{tab:e1_main}, different utility thresholds provide different operating points. At $\tau_\rho=0.9$, the utility-aware policy gives the most accuracy-oriented configuration in the current study. Compared with full reception, it reduces the average decision delay by $724.08$ ms and the average number of received packet blocks by $144.81$, while maintaining a matched-ground-truth rate of $0.8133$ compared with $0.8464$ under full reception. At $\tau_\rho=0.8$, the utility-aware policy provides a balanced operating point. It saves $1209.17$ ms and $241.83$ packet blocks relative to full reception, while maintaining a matched-ground-truth rate of $0.7741$. This operating point also outperforms the stability baseline in both task quality and stopping efficiency, improving the matched-ground-truth rate from $0.6687$ to $0.7741$ and reducing the average decision delay from $2368.36$ ms to $2322.29$ ms. At $\tau_\rho=0.7$, the policy provides a more aggressive early-stopping configuration when lower latency is prioritized over decision fidelity.

\subsection{Impact of Packet Arrival Order}

Packet-arrival order has a significant impact on progressive emergency decision. To evaluate this effect, utility-aware stopping with $\tau_\rho=0.8$ is tested under three packet-arrival orders: center-first, raster, and random. As shown in Table~\ref{tab:arrival}, the three schedules consume comparable packet budgets but lead to markedly different decision quality and delay.

Center-first delivery achieves a match rate of $0.7741$ with an average decision delay of $2322.29$ ms, whereas raster delivery decreases to $0.7681$ at $2693.60$ ms and random delivery further decreases to $0.4428$ at $2830.65$ ms. Compared with random delivery, center-first delivery improves the match rate by $33.13$ percentage points and reduces the average decision delay by $508.36$ ms. The main difference is therefore not the number of received packet blocks, but the order in which task-relevant regions become available. Center-first delivery exposes task-critical regions earlier, whereas random delivery leads to stronger scene fragmentation and delays reliable hazard recognition.

\begin{table}[h]
\caption{Impact of packet-arrival order under utility-aware stopping ($\tau_\rho=0.8$).}
\label{tab:arrival}
\centering
\begin{tabular}{lccc}
\toprule
Arrival Order & Match Rate & Blocks & Delay (ms) \\
\midrule
Center-first & 0.7741 & 464.46 & 2322.29 \\
Raster & 0.7681 & 538.72 & 2693.60 \\
Random & 0.4428 & 566.13 & 2830.65 \\
\bottomrule
\end{tabular}
\end{table}

This result should be interpreted from the packet-arrival perspective. In the present work, the arrival order is externally given, and the utility predictor is used only for receiver-side stopping rather than for packet scheduling. A natural implication is that the proposed stopping framework can be combined in future work with semantic packet scheduling or priority queuing.

\subsection{Robustness under Packet Loss}

\begin{figure}[t]
\centering
\includegraphics[width=0.95\columnwidth]{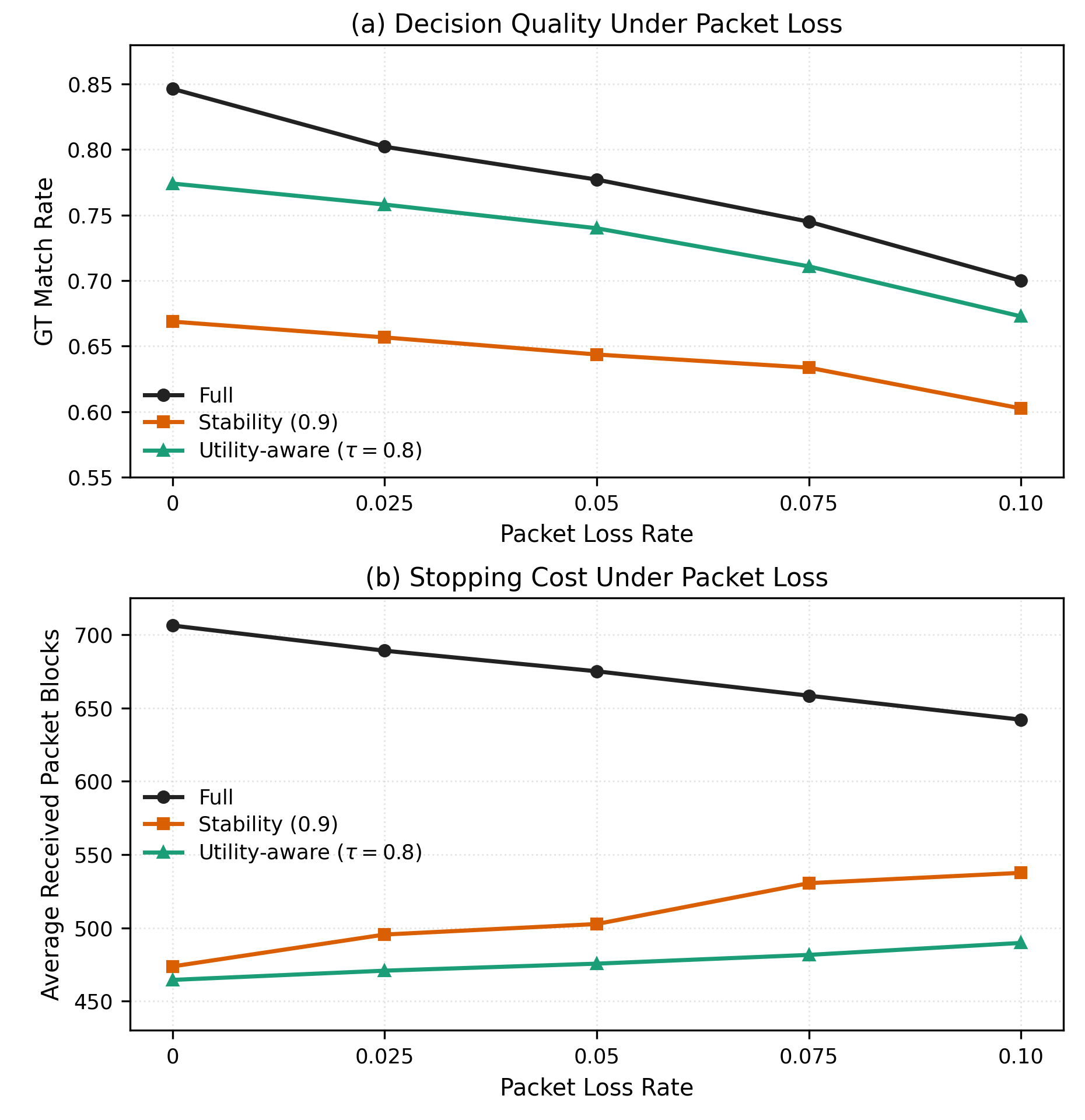}
\caption{Effect of packet loss on progressive machine decision. The top subfigure reports match rate, and the bottom subfigure reports average received packet blocks. Results are averaged over three random packet-loss seeds. Under increasing packet loss, the proposed utility-aware method remains closer to full reception than the stability baseline while requiring fewer received blocks.}
\label{fig:loss}
\end{figure}

Packet loss in practical emergency links imposes a dual penalty, because both decision quality and time to action may be degraded. To evaluate this effect under more realistic transmission conditions, the three main policies are tested under packet loss rates of $0\%$, $2.5\%$, $5\%$, $7.5\%$, and $10\%$, while the center-first arrival schedule is kept unchanged. Results are averaged over three random packet-loss seeds. As summarized in Fig.~\ref{fig:loss}, utility-aware stopping remains closer to the full-reception reference than the stability baseline under moderate packet loss.

More specifically, the match rate of full reception decreases from $0.8464$ at $0\%$ loss to $0.8022$, $0.7771$, $0.7450$, and $0.6998$ at $2.5\%$, $5\%$, $7.5\%$, and $10\%$ loss, respectively. The utility-aware policy follows a similar degradation trend, decreasing from $0.7741$ to $0.7580$, $0.7400$, $0.7108$, and $0.6727$, while remaining consistently closer to the full-reception reference than the stability baseline. By comparison, the stability baseline decreases from $0.6687$ to $0.6566$, $0.6436$, $0.6335$, and $0.6024$ over the same loss range. These results indicate that packet-level utility remains informative for receiver-side stopping even when packet delivery is degraded.

The communication-cost results show a consistent advantage for utility-aware stopping. As the loss rate increases from $0\%$ to $10\%$, the average number of received packet blocks rises from $464.46$ to $489.70$ for the utility-aware policy, compared with an increase from $473.67$ to $537.52$ for the stability baseline. The corresponding average decision delay increases from $2322.29$ ms to $2690.04$ ms for the utility-aware policy and from $2368.36$ ms to $2956.46$ ms for the stability baseline. Taken together, these results show that under packet loss, utility-aware stopping maintains higher task quality than the stability baseline while reaching a decision with fewer received packet blocks.

\section{Conclusion}

This paper presented a utility-aware progressive decision framework for emergency visual inference over packet streams. By quantifying packet-level task utility and accumulating it at the receiver, early stopping was enabled under partial reception and a given packet arrival order. Experimental results on the fire-scene detection dataset showed that the proposed method reduced decision delay and communication cost while preserving task-level decision quality. The results also showed that utility-aware stopping maintained its advantage over the stability-based baseline under moderate packet loss and different packet-arrival orders. These findings suggest that packet-level task utility can provide an effective basis for communication-efficient early stopping in emergency machine decision. Future work may extend the present framework to broader emergency scenarios, additional hazard categories, and different task backbones, as well as semantic packet scheduling and more general task-oriented transport design.

\section*{Acknowledgment}

This work was supported by the National Key Research and Development Program of China (No. 2024YFE0200302).

\bibliographystyle{IEEEtran}
\bibliography{globecom_udp_progressive_refs}

\end{document}